\documentclass[12pt]{article}

\usepackage{latexsym,amsmath,amssymb,theorem,epsfig}

\DeclareFontFamily{OT1}{rsfs10}{}
\DeclareFontShape{OT1}{rsfs10}{m}{n}{ <-> rsfs10 }{}
\DeclareMathAlphabet{\mathscript}{OT1}{rsfs10}{m}{n}

\numberwithin{equation}{section}

\newcommand{\ns}{\normalsize}

\def\g{\gamma}

\def\v{\varphi}

\def\gsim{ \lower .75ex \hbox{$\sim$} \llap{\raise .27ex \hbox{$>$}} }
\def\lsim{ \lower .75ex \hbox{$\sim$} \llap{\raise .27ex \hbox{$<$}} }
\def\be{\begin{equation}}
\def\ee{\end{equation}}
\def\bea{\begin{eqnarray}}
\def\eea{\end{eqnarray}}

\theoremstyle{plain}

{\theorembodyfont{\rmfamily} }

\begin{document}


\begin{titlepage}

\vspace{-5cm}

\title{
  \hfill{\ns }  \\[1em]
   {\LARGE Energy-Spin Trajectories in $AdS_5 \times S^5$ from Semiclassical
  Vertex Operators}
\\[1em] }
\author{
   Evgeny I. Buchbinder
     \\[0.5em]
   {\ns The Blackett Laboratory, Imperial College, London SW7 2AZ, U.K.}} 

\date{}

\maketitle

\begin{abstract}
We study the relation between vertex operators in $AdS_5 \times S^5$
and classical spinning string solutions. 
In the limit of large quantum numbers the 
treatment of vertex operators becomes
semiclassical. 
In this regime, a given vertex
operator carrying a certain set of quantum numbers defines a singular
solution. We show in a number of examples that this solution coincides
with the classical string solution with the same quantum numbers
but written in a different two-dimensional coordinate system. 
The marginality condition
imposed on an operator yields a relation between the energy 
and the other quantum numbers which is shown to coincide
with that
of the corresponding classical string solution.
We also argue that in some cases vertex operators in $AdS_5 \times S^5$
cannot be given by expressions similar to the ones in flat space
and a more involved consideration is required.

\end{abstract}

\thispagestyle{empty}

\end{titlepage}

\section{Introduction}

String theory on the $AdS_5 \times S^5$ background is considered 
to be integrable. In was shown in~\cite{BRP} that the 
equations of motion of the 
classical string sigma model on $AdS_5 \times S^5$
are integrable. However, it is also believed that this 
theory posses integrability at the quantum level. See the 
review article~\cite{AF} and references therein for the recent 
progress in this direction. A big evidence in favor of integrability 
comes from the field theory side of the AdS/CFT 
duality~\cite{Juan, Gubser, Witten}. First, computations 
of the anomalous dimensions of infinite length single trace
operators (corresponding to string states with infinite energy)
were summarized in~\cite{Beisert1, Beisert2} in the form 
of the asymptotic Bethe ansatz. Furthermore, it was recently proposed 
in~\cite{G1, F, G2, AFint} how to take into account the corrections
due to the finite length of these operators. 

Integrability suggests that it should be possible to 
identify all quantum  
states in $AdS_5 \times S^5$
(analogues of particles in flat space) 
as well as the vertex operators to describe 
their interaction. In fact, by AdS/CFT correspondence,
each vertex 
operator should be associated with 
a local gauge-invariant operator in ${\cal N}=4$
gauge theory in the planar limit~\cite{Polyakov}.
Furthermore, integrability also implies that it should also be possible 
to systematically compute the energy of all 
these string states at any t'Hooft coupling. 
At the present moment, 
it is not known how to perform this program.
In particular, it is not much known about 
construction of
the vertex operators in this theory.

This paper is devoted to studying the vertex operators in $AdS_5 \times S^5$
in the limit of large quantum numbers where the analysis becomes
semiclassical.  Vertex operators in the semiclassical approximation 
were previously studied in~\cite{Gubser, Polyakov, Tseytlinver}.
Our aim is to find the relation between them 
and various
classical spinning string solutions.
For a review of string solutions 
see~\cite{Tseytlinrev} and references therein. 
A priory, it is not very clear why such
a relation should exist. A vertex operator defines a particle-like quantum
state with certain energy, spins and additional quantum numbers.
This is not in an obvious way related to any classical
solution. 
However, for large quantum numbers, quantum states
often can be approximated with classical field trajectories.
Moreover, a semiclassical computation of correlation functions
of vertex operators is equivalent to finding a certain classical
solution with singularities at the positions of the operators.
So in the semiclassical regime both descriptions involve classical solutions.
Let us say, we start with a classical string solution with a set
of conserved quantities (energy, spins, angular momenta in
various directions, $\ldots$). If we construct a vertex operators
carrying the same quantum numbers, we should expect that 
in the limit of large charges such an operator corresponds 
to this classical solution.
In fact, we will argue and show it explicitly
in a number of examples that if we compute the two-point
function of these vertex operators we obtain the same
classical solution we started with but written in a different 
two-dimensional coordinate
system. The two coordinate systems are related by a conformal
transformation which is singular at the points where operators are
inserted. 
In this paper, we will present a several examples of this
correspondence. We will also show that the relation between 
the energy and the remaining quantum numbers 
of the vertex operators coincides with the similar relation
of the corresponding classical solutions. Note that since construction
of vertex operators in $AdS_5 \times S^5$ is not well understood,
one can also view our procedure as a consistency check on the 
vertex operators themselves.

The paper is organized as follows.
In Section 2, we will explain how exactly the singular solution
obtained from inserting vertex operators is related to the corresponding
classical string solution. For this we will consider the correlation
function of two vertex operators in the limit of large quantum numbers
so that the description becomes semiclassical. We explain why the
semiclassical trajectory in the presence of the operators should coincide
with the classical solution carrying the same charges in a different
two-dimensional coordinate system. In this framework, the 
relation between the
energy and the remaining quantum numbers comes from requiring that the
vertex operator be of the right conformal dimension.
As the first demonstration of this approach
we consider an example of a string spinning in a two-dimensional
plane in flat space~\cite{Tseytlinver}.

In the rest of the paper, we perform this analysis for various string
solutions in $AdS_5 \times S^5$. 
More precisely, we will consider strings spinning in $AdS_3$ and $S^3$.
In these cases the equations of motion are 
non-linear. We will solve them by starting with a classical solution 
and performing the appropriate conformal transformation. 
In Section 3, we consider a
folded string spinning in $AdS_3$~\cite{GKP} and derive the logarithmic
correction to the energy. In Section 4, we consider a spinning string
in $S^3$. This case is subtle since there are two different solutions
with the same energy and spin. So an important question is how to
distinguish these two states with vertex operators.
One of these solutions has a flat space limit
and, thus, its vertex operator can be written by analogy with
vertex operators in flat space. However, the second solution does not
have such a limit. Furthermore, it carries an additional quantum number
which is a topologically trivial winding. We propose
that the vertex operator for this state should be written in terms of
the T-dual variables. In general, T-duality on $AdS_5 \times S^5$ is
rather non-trivial~\cite{Ricci1, BM, Ricci2}. However, in our case the
semiclassical calculations evade this problem since most of the $S^5$
coordinates are constants. We compute the energy-angular momentum trajectory
of this operator and show that it coincides with that
of the corresponding classical solution. 
This can be viewed
as a consistency check on the proposed vertex operator.


\section{Vertex Operators and Classical String Solution}


In this section, we would like to describe the relation between
semiclassical vertex operators and classical string solutions.
We will be interested in computing the correlation function
of two such operators. These operators are inserted on the complex plane
whose coordinates we will denote by $(\xi, \bar \xi)$. On the other
hand the classical closed string world-sheet is the cylinder
whose coordinates we will denote by $(\tau, \sigma)$ and that is where
classical string solutions are defined. We will perform a Wick rotation
$\tau=-i \tau_e$ and consider the Euclidean  world-sheet.
Now we perform a conformal transformation to map the cylinder to the
plane (or, more precisely, to the sphere)
\begin{equation}
z=e^{\tau_e + i \sigma}\,.
\label{1.1}
\end{equation}
In doing this, the points $\tau_e =\pm \infty$ are mapped to $z=0$ and
$z=\infty$. Now we perform one more conformal transformation to map the
point $z=\infty$ to a finite position. From $(z, \bar z)$ we go to
$(\xi, \bar \xi)$
\begin{equation}
z=\frac{\xi}{\xi-\xi_1}\,.
\label{1.2}
\end{equation}
Now $z=0$ goes to $\xi=0$ and $z=\infty$ goes to $\xi=\xi_1$.
Thus, this sequence of conformal transformations creates two
singularities on the $\xi$-plane, one at $\xi=0$ and the other one
at $\xi=\xi_1$. We will view these two points as the points where
two vertex operators are inserted. Of course, we can perform the second
conformal transformation to move the singularity at $\xi=0$ to
an arbitrary point $\xi_2$,
\be
z=\frac{\xi-\xi_2}{\xi-\xi_1}\,.
\label{1.2.1}
\ee
However, by translational invariance
we can move the position of one of the operators to $\xi=0$.
Hence, for simplicity, we will set $\xi_2=0$.
If we start with a classical solution on the cylinder
and perform the conformal transformations~\eqref{1.1} and~\eqref{1.2}
we obtain a solution on the $\xi$-plane with singularities
at $\xi=0$ and $\xi=\xi_1$. The singularities arise due to
specifics of these conformal transformations. This singular solution
is what we expect to find if we insert the appropriate semiclassical
vertex operators at $\xi=0$ and $\xi=\xi_1$. One of our goals
in the rest of the paper will be to check this statement in various examples.
Once the solution is found, we can calculate semiclassically
the two-point function by evaluating the action on this solution.
Conformal invariance requires that the dimension of the operator
be $(1, 1)$. On the other hand, in general, the expected behavior
of the two-point
function is $|\xi_1|^{2 \gamma -4}$, where $\gamma$ can be viewed
as the anomalous dimension. For operators of dimension $(1, 1)$
the anomalous dimension $\gamma$ vanishes.
Setting $\gamma=0$ is supposed to yield the
dependence of the energy on the other quantum numbers.
Note that in finding the semiclassical solution, all parameters must be
fixed in terms of the quantum numbers carried by the vertex operator.
Otherwise, if there is an arbitrary parameter 
which is not fixed, 
setting $\gamma$ to zero will yield the energy as a function
not just of the remaining quantum numbers but also on
this unfixed parameter.

Let us finish our general discussion with two simple remarks.
The first remark concerns when the semiclassical analysis is reliable.
Clearly, to ignore quantum corrections we have to take
$\alpha^{\prime}$ to be small or the t'Hooft coupling
$\lambda=1/\alpha^{\prime 2}$ to be large. Furthermore, we have
to take the energy and other quantum numbers to be large.
The second remark is that in our semiclassical computations we
can ignore the dependence of the vertex operators on the fermions
and take them purely bosonic.

The way we are going to proceed in this paper is as follows.
We will start with a specific classical string solution. Then we will
write the vertex operator carrying the same quantum numbers
as the classical solution. The construction of vertex operators
in $AdS_5 \times S^5$ is not well understood and we will have to make
certain reasonable guesses. For example, in many cases, the classical
string solution admits a flat space limit when the string
becomes very small and does not feel the curvature of
$AdS_5$ and $S^5$. In these cases, the corresponding vertex
operator is expected to have the similar form as in flat space.
Once we write the vertex operator, we proceed with
the semiclassical evaluation of the two-point function
and the computation of the energy-spin relation. In all examples below,
this relation will coincide with that of the classical
solution we started with.

Let us first illustrate the above procedure in the example
of a string spinning in the plane in flat space
following~\cite{Tseytlinver}.
The relevant classical action is
\begin{equation}
A=-\frac{\sqrt{\lambda}}{4 \pi}\int d \tau d \sigma
(-\partial_a t \partial^a t + \partial_a X \partial^a \bar X)\,.
\label{1.3}
\end{equation}
The classical solution of interest is
\be
t =\kappa \tau \,,{\ } X= \omega \sin \sigma e^{i \tau}\,, {\ }
\bar X= \omega \sin \sigma e^{-i \tau}\,.
\label{1.3.1}
\end{equation}
Furthermore, the Virasoro constraints give $\omega =\kappa$.
This solution has two conserved quantities, the energy $E$ and
the spin $S$. They are given by
\be
E=\frac{\sqrt{\lambda}}{2 \pi} \int_{0}^{2 \pi} d \sigma
\dot{t}= \sqrt{\lambda} \kappa
\label{1.4}
\end{equation}
and
\be
S= \frac{i \sqrt{\lambda}}{4 \pi} \int_{0}^{2 \pi} d \sigma
(X \dot{\bar X}- \bar X \dot{X}) =\frac{ \sqrt{\lambda} \omega^2}{2}\,.
\label{1.5}
\ee
Note that both $E$ and $S$ scale as $\sqrt{\lambda}$. Since $\omega =\kappa$
we get the following relation between $E$ and $S$
\be
E=\sqrt{2 \sqrt{\lambda} S}=\sqrt{\frac{2}{\alpha^{\prime}} S}\,.
\label{1.6}
\ee
Now we will consider the vertex operator carrying the energy and the
spin in the $(X, \bar X)$-plane. In flat space it is given by
\be
V_S= e^{-i E t} (\partial X {\bar \partial} X)^{S/2}\,.
\label{1.7}
\ee
We want to compute the correlation function of $V_S$
inserted at $\xi=0$ and $V_{-S}$,
\be
V_{-S}=e^{i E t} (\partial {\bar X} {\bar \partial} \bar X)^{S/2}\,,
\label{1.8}
\ee
inserted at $\xi=\xi_1$ in the limit of large $\sqrt{\lambda}, E, S$.
We perform a Euclidean rotation
\be
\tau=-i \tau_e\,, \, t=-i t_e\,,\, A=i A_e
\label{1.8.1}
\end{equation}
and consider the action on the plane modified by the
vertex operators~\eqref{1.7} and~\eqref{1.8}
(to simplify our notation we will still denote it by $A_e$)
\bea
A_e & = &\frac{\sqrt{\lambda}}{\pi} \int d^2 \xi
(\partial t_e \bar \partial t_e +
\frac{1}{2} \partial X \bar \partial \bar X +\frac{1}{2}
\partial \bar X  \bar \partial X) \nonumber \\
 & + & E \int d^2 \xi t_e (\delta^2 (\xi)-\delta^2 (\xi-\xi_1))
\nonumber \\
&- &\frac{S}{2} \int d^2 \xi \delta^2 (\xi)
\ln (\partial X \bar \partial X)
- \frac{S}{2} \int d^2 \xi \delta^2 (\xi-\xi_1)
\ln (\partial \bar X \bar \partial \bar X)\,,
\label{1.9}
\eea
where $d^2\xi= d Re(\xi) d Im(\xi)$.
The two-point function in the semiclassical approximation
is then simply given by
\be
\langle V_S(0) V_{-S}(\xi_1)\rangle \sim e^{-A_e}\,.
\label{1.9.1}
\end{equation}
Here $A_e$ is evaluated on the solution to the equations
of motion which are as follows
\bea
&& (\partial \bar \partial + \bar \partial \partial) t_e
= \frac{\pi E}{\sqrt{\lambda}} (\delta^2 (\xi)-\delta^2 (\xi-\xi_1))\,,
\label{1.10}
\\
&&(\partial \bar \partial + \bar \partial \partial) \bar X=
\frac{\pi S}{\sqrt{\lambda}}
\left[ \partial\left( \frac{\delta^2 (\xi)}{\partial X}\right) +
\bar \partial\left( \frac{\delta^2 (\xi)}{\bar \partial X}\right)\right]\,,
\label{1.11}
\\
&&(\partial \bar \partial + \bar \partial \partial) X=
\frac{\pi S}{\sqrt{\lambda}}
\left[ \partial\left( \frac{\delta^2 (\xi-\xi_1)}
{\partial \bar X}\right) +
\bar \partial\left( \frac{\delta^2 (\xi-\xi_1)}
{\bar \partial \bar X}\right)\right]\,.
\label{1.12}
\eea
According to our general discussion, the solution
to this system is given by the classical solution~\eqref{1.4}
written in the coordinates $(\xi, \bar \xi)$.
More precisely
\bea
&&t_e =\frac{\kappa}{2}[\ln \xi \bar \xi - \ln (\xi-\xi_1)
(\bar \xi-\bar \xi_1)]\,,
\nonumber
\\
&& X=\frac{\omega}{2 i}\left( \frac{\xi}{\xi-\xi_1}-
 \frac{\bar\xi }{\bar \xi-\bar\xi_1}\right)\,,
\nonumber
\\
&& \bar X=\frac{\omega}{2 i}
\left( \frac{\bar \xi-\bar \xi_1}{\bar\xi }-
 \frac{\xi -\xi_1}{\xi }\right)\,.
\label{1.14}
\eea
Substituting~\eqref{1.14} into eqs.~\eqref{1.10}-\eqref{1.12}
we find that it is indeed a solution\footnote{To prove that it is
a solution we have to use the well-known relations like
$\bar \partial \frac{1}{\xi}= \pi \delta^2(\xi)$.}
with parameters $\kappa$ and
$\omega$ fixed in terms of the quantum numbers of the operator
\be
\kappa=\frac{E}{\sqrt{\lambda}}\,, \quad \omega^2=\frac{S}{\sqrt{\lambda}}\,.
\label{1.15}
\ee
The two-point function $\langle V_S(0) V_{-S}(\xi_1)\rangle$
is given by the action evaluated on the solution~\eqref{1.14}.
Ignoring the obvious divergence $\sim \ln(0)$, we obtain
\be
\langle V_S(0) V_{-S}(\xi_1)\rangle \sim
|\xi_1|^{E^2/\sqrt{\lambda}-2 S}\,.
\label{1.16}
\ee
In other words, we get
\be
2 \gamma-4 = \frac{E^2}{\sqrt{\lambda}}-2 S\,.
\label{1.16.1}
\ee
Remembering that our analysis is valid only in the regime of large
$E$ and $S$ 
(so that $-4$ can be ignored)~\footnote{More precisely, we have to take the limit 
of large $E$, $S$, $\lambda$ so that $E$, $S \sim \sqrt{\lambda}$.}
we find that the condition $\gamma=0$
yields
\be
E=\sqrt{2 \sqrt{\lambda} S}=\sqrt{\frac{2}{\alpha^{\prime}} S}\,.
\label{1.17}
\ee
This is the same energy-spin relation as for the corresponding
classical string solution.
Note that eq.~\eqref{1.17} is the standard Regge trajectory
for string states in flat space in the limit of large $E$ and $S$.

In the rest of the paper, we apply this consideration
for various states in $AdS_5 \times S^5$. The equations of motion
in these cases are very non-linear and, a priory, it is not clear
how to solve them. However, as was explained above we can construct
the solutions starting with the corresponding classical string solution
on the cylinder
and performing the conformal transformations~\eqref{1.1} and~\eqref{1.2}.


\section{Folded Strings in $AdS_3$}


As the first example of strings in $AdS_5 \times S^5$ we will
consider a folded string spinning in $AdS_3$~\cite{GKP}. In
field theory, this string state corresponds to the twist two
operators. We will describe $AdS_5$ in global coordinates in
which the metric takes the form
\begin{equation}
d s^2 = d \rho^2 -\cosh^2\rho d t^2 +
\sinh^2\rho (d \theta^2 + \cos^2 \theta d \phi_1^2 + \sin^2\theta
d \phi_2^2)\,.
\label{2.1}
\end{equation}
The relation to the embedding coordinates in $R^{2, 4}$ is as
follows
\bea 
&& Y_5 + i Y_0 =\cosh\rho {\ }e^{ i t}\,, \quad Y_1+ i Y_2
=\sinh \rho \cos \theta {\ }e^{ i \phi_1}\,, 
\nonumber \\ 
&& Y_3+ i Y_4 =\sinh \rho \sin \theta {\ }e^{ i \phi_2} \,, 
\nonumber \\
&&-Y_0^2-Y_5^2+Y_1^2+Y_2^2+Y_3^2+Y_4^2=-1\,. 
\label{2.2}
\end{eqnarray}
Since the string is spinning in $AdS_3$ we can set
\be 
\theta =0\,, \quad \phi_2=0 
\label{2.2.1}
\end{equation}
and denote
\bea 
&& Y=\cosh\rho {\ }e^{i t}\,, 
\quad \bar Y= \cosh\rho {\ }e^{-i t}\,, \nonumber \\
&& X=\sinh \rho {\ }e^{i \phi_1}\,, \quad \bar X= 
\sinh\rho {\ }e^{-i\phi_1}\,. 
\label{2.3} 
\end{eqnarray}
The action is then given by
\begin{equation} 
A=-\frac{\sqrt{\lambda}}{4 \pi} \int d^2 \sigma
(-\cosh^2 \rho {\ }\partial_a t \partial^a t+\partial_a \rho \partial^a
\rho + \sinh^2 \rho {\ }\partial_a \phi_1 \partial^a \phi_1)\,. 
\label{2.3.1}
\end{equation}
The folded string solution in the limit of large 
$E$ and $S$ has the following structure~\cite{GKP}
\begin{equation}
t=\kappa \tau\,, \quad \phi_1= \omega \tau\,, \quad \rho= \mu \sigma \,.
\label{2.4}
\end{equation}
\begin{figure}[ht]
\centering
\includegraphics[width=100mm]{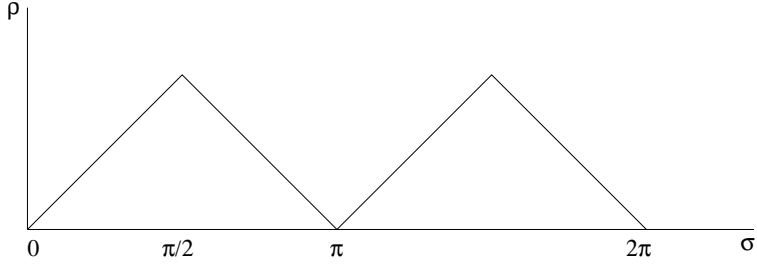}
\caption{The approximate solution $\rho=\rho(\sigma)$ in the 
limit of large energy and spin.}
\label{fig1}
\end{figure}
The equation of motion for $\rho$ is satisfied only 
if~\footnote{This relation is valid only in the limit of large
$S$.} 
\be
\kappa=\omega\,.
\label{eqro}
\ee
Furthermore, the equations of motion must be supplemented
with the Virasoro constraints. If we denote by $X_M$, 
$M=1, \dots, 6$ the coordinates on $S^5$ they read 
\be
\eta^{P Q} (\dot{Y}_P \dot{Y}_Q+ Y_P^{\prime} Y_Q^{\prime})+
\dot{X}_M\dot{X}_M +X_M^{\prime} X_M^{\prime}=0
\label{Vir1}
\ee
and 
\be
\eta^{P Q} \dot{Y}_P Y_Q^{\prime}+
\dot{X}_M X_M^{\prime}=0\,,
\label{Vir2}
\ee
where $\eta_{P Q}$ is the metric in $R^{2, 4}$,
\be
\eta_{P Q}={\rm diag} (-1, -1, 1, 1, 1, 1)\,.
\label{2.4.1}
\ee
In our case the Virasoro constraints yield
\begin{equation}
\kappa= \mu\,.
\label{2.5}
\ee
Note that $\rho$ has to be a periodic function of $\sigma$. Thus,
this solution cannot be valid on the whole cylinder $\sigma \in [0,
2 \pi]$. In fact, it is valid only on the interval $\sigma \in [0,
\pi/2]$. Then it has to be folded and continued by periodicity. The
form of the solution in the limit of large energy and spin is
depicted in~Figure 1. This solution is not smooth and is only 
approximate in the limit of large $E$ and $S$.
The exact solution is also possible to find and
it is given in terms of elliptic integrals. We will not need the
details of the exact solution in this paper. They can be found
in~\cite{GKP, FTsemi, Tseytlinrev}.

The energy and the spin of this solution are given by
\be
E= \sqrt{\lambda} \int_0^{2 \pi} \frac{d \sigma}{2 \pi}
\cosh^2\rho {\ }\dot{t} =\frac{\sqrt{\lambda}}{2 \pi} \kappa \int_0^{2
\pi} \frac{d \sigma}{2 \pi} \cosh^2 \rho
\label{2.6}
\end{equation}
and
\be
S= \sqrt{\lambda} \int_0^{2 \pi} \frac{d \sigma}{2\pi} \sinh\rho
{\ }\dot{\phi_1} =\sqrt{\lambda} \kappa \int_0^{2 \pi} \frac{d \sigma}{2
\pi} \sinh^2 \rho \,.
\label{2.7}
\end{equation}
From these equations one can conclude that $\kappa \sim \ln S$. Therefore,
in the limit of large $E$ and $S$ the size of the string $\rho$
becomes very large. In this limit one can eliminate $\kappa$ in
eqs.~\eqref{2.6} and~\eqref{2.7} to find that $E$ and $S$ are
related as follows
\be E=S+ \frac{\sqrt{\lambda}}{\pi} \ln \frac{S}{\sqrt{\lambda}}
+\dots \,, \label{2.8} \ee
where the ellipsis stands for the subleading correction. The aim
of this section is to reproduce this dependence using
semiclassical vertex operators. One can find the form of the vertex
operator in this case by noticing that in the limit of small $\rho$
the string becomes very small and the curvature of $AdS_3$ can be
neglected. Hence, in this limit we have a string spinning in the
plane in flat space. In this limit the vertex operator should
coincide with~\eqref{1.7}. Thus, the natural expression for the
vertex operator for a string spinning in $AdS_3$ is
\begin{eqnarray} && V_S = Y^{-E} (\partial X \bar \partial
X)^{S/2}\,, \nonumber \\ && V_{-S}={\bar Y}^{-E} (\partial \bar X
\bar \partial \bar X)^{S/2}\,, \label{2.9} \eea
where $E$ is the energy and $S$ is the spin of the corresponding
state and $X$ and $Y$ are now given by eq.~\eqref{2.3}. We want to
compute the two-point function of the operator $V_S$ inserted on the
complex plane at $\xi=0$ and the operator $V_{-S}$ inserted at
$\xi=\xi_1$. The Euclidean action on the plane (including the
operator insertions) is as follows
\bea 
&& A_e=\frac{\sqrt{\lambda}}{\pi} \int d^2 \xi (\partial\rho
\bar \partial \rho+ \cosh^2 \rho {\ }\partial t_e \bar \partial t_e +
\sinh^2 \rho {\ }\partial \phi_1 \bar \partial \phi_1) 
\nonumber \\ 
&&+E
\int d^2 \xi \ln Y (\delta^2 (\xi)-\delta^2(\xi-\xi_1)) 
\nonumber \\
&& -\frac{S}{2} \int d^2 \xi \delta^2 (\xi) \ln (\partial X \bar
\partial X) - \frac{S}{2} \int d^2 \xi \delta^2 (\xi-\xi_1) \ln
(\partial \bar X \bar \partial \bar X)\,, 
\label{2.10} 
\eea
where we have also Wick rotated the $AdS$ time $t$.  
As we mentioned above, in the regime of large $E$ and $S$ (in which
our semiclassical analysis is valid) $\rho$ becomes very large. In
this limit $\cosh \rho \sim \sinh \rho \sim e^{\rho}$. 
%
%
Then the action becomes
\bea
&& A_e=\frac{\sqrt{\lambda}}{\pi} \int d^2 \xi (\partial\rho \bar
\partial \rho+ \cosh^2 \rho {\ }\partial t_e \bar \partial t_e + \sinh^2
\rho {\ }\partial \phi_1 \bar \partial \phi_1) \nonumber \\
&&+E \int d^2 \xi  (\delta^2 (\xi)-\delta^2(\xi-\xi_1)) t_e -S \int
d^2 \xi (\delta^2 (\xi)-\delta^2(\xi-\xi_1)) i \phi_1 \nonumber \\ &&
+ (E-S) \int d^2 \xi (\delta^2(\xi)+\delta^2(\xi-\xi_1)) \rho
\nonumber\\
&& -\frac{S}{2} \int d^2 \xi \delta^2 (\xi) \ln [\partial(\rho + i
\phi_1)] -\frac{S}{2} \int d^2 \xi \delta^2 (\xi) \ln
[\bar\partial(\rho + i \phi_1)] \nonumber \\
&& -\frac{S}{2} \int d^2 \xi \delta^2 (\xi-\xi_1) \ln [\partial(\rho
- i \phi_1)] -\frac{S}{2} \int d^2 \xi \delta^2 (\xi-\xi_1) \ln
[\bar\partial(\rho - i \phi_1)]\,.
\label{2.12}
\eea
We will start with the equation of motion for $\rho$,
\begin{eqnarray}
&& \partial \bar \partial \rho- \sinh \rho \cosh \rho (\partial t_e
\bar \partial t_e +\partial \phi_1 \bar \partial \phi_1)= \frac{\pi
(E-S)}{2 \sqrt{\lambda}}(\delta^2 (\xi)+\delta^2 (\xi-\xi_1))
\nonumber \\
&& +\frac{\pi S}{4 \sqrt{\lambda}}\left[ \partial \left(
\frac{\delta^2(\xi)}{\partial(\rho+i \phi_1)}\right) +\partial \left(
\frac{\delta^2(\xi-\xi_1)}{\partial(\rho-i \phi_1)}\right) +\partial
\to \bar \partial\right]\,.
\label{2.13}
\end{eqnarray}
According to our prescription, we expect the solution to be of the
form~\eqref{2.4} written in the coordinates $(\xi, \bar \xi)$. In
particular, we expect $\phi_1$ to behave as
\be
i \phi_1 =\omega \tau_e =\frac{\omega}{2} [\ln \xi \bar \xi -\ln
(\xi-\xi_1)(\bar \xi-\bar \xi_1)]\,.
\label{2.14}
\end{equation}
So
\be
\partial (i \phi_1) \sim \frac{1}{\xi (\xi-\xi_1)}\,. \label{2.14.1}
\end{equation}
This implies that
\be
\frac{\delta^2(\xi)}{\partial(\rho+i \phi_1)} \sim \xi (\xi-\xi_1)
\delta^2 (\xi)=0\,, \quad \frac{\delta^2(\xi-\xi_1)}{\partial(\rho+i
\phi_1)} \sim \xi (\xi-\xi_1) \delta^2 (\xi-\xi_1)=0\,.
\label{2.14.2} \ee
Thus, all terms in the second line vanish. Furthermore, we similarly
expect that
\be
t_e =\frac{\kappa}{2}[\ln \xi \bar \xi -\ln (\xi-\xi_1)(\bar
\xi-\bar \xi_1)]\,.
\label{2.15} \ee
Therefore,
\be
\partial t_e \bar\partial t_e+\partial \phi_1 \bar\partial \phi_1 \sim
\frac{(\kappa^2-\omega^2)}{|\xi||\xi-\xi_1|}\,.
\label{2.16}
\end{equation}
If we multiply both sides of~\eqref{2.13} by $|\xi||\xi-\xi_1|$ we
find that the delta-functional term vanishes and the equation for
$\rho$ becomes non-singular as is expected since the classical solution
for $\rho$ does not develop singularities under the conformal
transformation to the $\xi$-plane. Since the equation is the same as
in the absence of the operators, we find the same solution as before.
In the limit of large parameters we get
\bea
&& t_e \approx i \phi_1 =\frac{\kappa}{2}[\ln \xi \bar \xi -\ln
(\xi-\xi_1)(\bar \xi-\bar \xi_1)]\,, \nonumber \\
&&\rho= \mu \sigma =\frac{\mu}{2}\left(\ln \frac{\xi}{\bar \xi}-
\ln \frac{\xi-\xi_1}{\bar \xi-\bar \xi_1} \right)\,, 
\quad \sigma \in [0, \pi/2]
\label{2.17} \eea
and $\rho$ is folded as is Figure 1. Note that so far $\kappa$ and
$\mu$ are unrelated. Now let us consider equations of motion for
$t_e$ and $\phi_1$. They read
\be
\partial (\cosh^2 \rho {\ }\bar \partial t_e)+
\bar \partial (\cosh^2 \rho{\ }
\partial t_e)=\frac{\pi E}{\sqrt{\lambda}}(\delta^2
(\xi)-\delta^2(\xi-\xi_1))
\label{2.19} \ee
and
\be
\partial (\sinh^2 \rho {\ }\bar \partial \phi_1)+\bar \partial (\sinh^2
\rho {\ }\partial \phi_1)=-\frac{i\pi S}{\sqrt{\lambda}}(\delta^2
(\xi)-\delta^2(\xi-\xi_1))\,.
\label{2.20} \ee
Note that the additional delta-functional terms of the type
$\frac{\delta^2(\xi)}{\partial(\rho+i \phi_1)}$ which also arise in
eq.~\eqref{2.20} vanish by the same arguments as given below eq.~\eqref{2.14}.
We have to show that the expressions
in~\eqref{2.17} indeed solve these two equations. First, we
subtract~\eqref{2.20} from~\eqref{2.19}. Since $t_e \approx i \phi_1$
we get
\be
(\partial \bar \partial + \bar \partial
\partial)t_e=\frac{E-S}{\sqrt{\lambda}}\pi (\delta^2
(\xi)-\delta^2(\xi-\xi_1))\,.
\label{2.21} \ee
We see that~\eqref{2.17} is indeed a solution if we set
\be \kappa =\frac{E-S}{\sqrt{\lambda}}\,.
\label{2.22} \ee
Now we consider eq.~\eqref{2.20}. At this point we can approximate
$\sinh \rho \sim e^{\rho}$
\be
\partial (e^{2\rho} \bar \partial \phi_1)+\bar \partial (e^{2\rho}
\partial \phi_1)=-\frac{i\pi S}{\sqrt{\lambda}}(\delta^2
(\xi)-\delta^2(\xi-\xi_1))\,.
\label{2.23} \ee
At the first glance it looks problematic that~\eqref{2.17} can be
a solution if $\rho$ is large and behaves as $\mu \sigma$. 
However, our conformal transformations~\eqref{1.1},~\eqref{1.2}
are well-defined only for function on the cylinder, that is 
periodic in $\sigma$. Thus, we
have to use the fact that $\rho$ behaves as $\mu \sigma$ only for $\sigma
\in [0, \pi/2]$ and really is a periodic function. Periodicity
of $\rho$ means that
we can expand it in the Fourier series
\be
e^{2 \rho}= (e^{2 \rho})_0 +\sum_{n\neq 0} (e^{2 \rho})_n e^{i n
\sigma}\,.
\label{2.24} \ee 
Now we substitute~\eqref{2.24} into~\eqref{2.23}. We know that away
from the singularities $\xi=0$, $\xi=\xi_1$, eq.~\eqref{2.23} is the
classical equation which is satisfied by~\eqref{2.17}. So all we
have to understand is where the singular terms in the left hand side
come from. It is straightforward to realize that singular
contributions on the left hand side can come only from the zero
Fourier component $(e^{2 \rho})_0$. All the higher Fourier modes do
not contribute to the delta-functions. The zero Fourier mode is easy
to find:
\be
(e^{2 \rho})_0= \int_0^{2 \pi} \frac{d \sigma}{2 \pi} e^{2 \rho} = 4
\int_0^{\pi/2}\frac{d \sigma}{2 \pi} e^{2 \mu \sigma} =
\frac{1}{\pi \mu} (e^{\pi \mu}-1)\,.
\label{2.25} \ee
If we now substitute~\eqref{2.25} into~\eqref{2.23} we find that the
logarithmic function $\phi$ in eq.~\eqref{2.17}
solves~\eqref{2.23} provided $\mu$ is related to the spin $S$ as
follows
\be
\mu = \frac{1}{\pi}\ln \frac{S}{\sqrt{\lambda}} + {\cal O}(1)\,.
\label{2.26} \ee
The $t_e$- and $\phi_1$- dependent terms in
the action are straightforward to compute. 
Up to the irrelevant divergence $\sim \ln (0)$
they give
\be
A_e (t_e, \phi)=-\sqrt{\lambda} \kappa^2 \ln |\xi_1|\,.
\label{2.28} \ee
The $\rho$-dependent part of the action,
\be
A_e (\rho) =\frac{\sqrt{\lambda}}{\pi} \int d^2 \xi \partial \rho
\bar \partial \rho\,,
\label{2.29} \ee
is easier to compute if we go back to the $(\tau_e, \sigma)$
coordinates. In these coordinates $\rho$ is just $\mu \sigma$ and,
hence, we obtain
\be
A_e (\rho)=\frac{\sqrt{\lambda}}{4 \pi} 2 \pi \mu^2
\int_{\tau_1}^{\tau_2} d \tau= \frac{\sqrt{\lambda} \mu^2}{2}
(\tau_2-\tau_1)\,,
\label{2.30} \ee
where $\tau_2 \to \infty$ and $\tau_1 \to -\infty$. If we now go to
the ${\xi, \bar \xi}$ coordinates we have
\be
\tau_2 =\frac{1}{2} \ln \left(\frac{\xi \bar \xi}{(\xi-\xi_1)(\bar
\xi-\bar\xi_1) } \right)|_{\xi\to \xi_1}
\label{2.32.1} \ee
and
\be
\tau_1 =\frac{1}{2} \ln \left(\frac{\xi \bar \xi}{(\xi-\xi_1)(\bar
\xi-\bar\xi_1) } \right)|_{\xi\to 0}\,.
\label{2.32.2} \ee
Substituting it into~\eqref{2.30} and ignoring the obvious divergence
$\sim \ln(0)$ we find that
\be
A_e (\rho)=\sqrt{\lambda}\mu^2 \ln |\xi_1|\,.
\label{2.33} \ee
Combining it with $A_e(t_e, \phi)$ we obtain
\be
\langle V_S(0) V_{-S}(\xi_1)\rangle \sim e^{-A_e}\sim
|\xi_1|^{\sqrt{\lambda} (\kappa^2-\mu^2)}\,.
\label{2.34} \ee
Thus, in the limit of large $\lambda$  
the marginality condition implies that
\be
\kappa=\mu \,.
\label{2.35} \ee
Recalling that $\kappa= (E-S)/\sqrt{\lambda}$ and $\mu =
\frac{1}{\pi}\ln \frac{S}{\sqrt{\lambda}}$ we obtain the following
relation between the energy and the spin
\be
E-S= \frac{\sqrt{\lambda}}{\pi} \ln \frac{S}{\sqrt{\lambda}} +{\rm
subleading} {\ } {\rm terms}  \,.
\label{2.36} \ee
Thus, we have obtained the same energy-spin dependence as for the
corresponding classical solution. As was already stated above, our
analysis is valid only in the regime of large $E$ and $S$ and, thus,
it cannot be used for computing the subleading terms in~\eqref{2.36}.
To compute the subleading corrections one has to go beyond the
semiclassical approximation and, in particular, include the
dependence on the fermions in the vertex operators. This is beyond
the scope of the present paper.


\section{Strings Spinning in $S^3$}


In the last section, we will consider solutions describing strings
spinning in $S^3 \subset S^5$. For given energy $E$ and angular
momentum $J$ there are two different solutions of this
type~\cite{FTmulti}. Thus, there is the question whether we can write
vertex operators distinguishing these two solutions.
Before we discuss the vertex operators, let us
first set up our notation. We will perform our analysis in global
coordinates in which the metric on $S^5$ looks as follows
\be
ds^2= d\gamma^2 +\cos^2 \gamma d \varphi_3^2 +\sin^2 \gamma(d
\psi^2+\cos^2\psi d \varphi_1^2 +\sin^2\psi d\varphi_2^2)\,.
\label{3.2} \ee
The embedding coordinates in $R^6$ can then be written as
\bea
&& X_1+i X_2 =\sin \gamma \cos \psi e^{i \v_1}\,, \quad X_3+i X_4
=\sin \gamma \sin \psi e^{i \v_2}\,, \nonumber \\
&& X_5 +i X_6 = \cos \gamma e^{i \v_3}\,, \nonumber \\
&& X_1^2+X_2^2+X_3^2+X_4^2+X_5^2+X_6^2=1\,.
\label{3.1} \eea
Since string is spinning in $S^3$ we will set for the rest of the
section
\be
\v_3=0\,.
\label{3.1.1} \ee
Both $S^3$-solutions are located at $\rho=0$ and the
only non-trivial dependence on the $AdS_5$ coordinates is
\be
t =\kappa \tau
\label{3.3} \ee
which yields the energy
\be
E=\sqrt{\lambda}\kappa\,.
\label{3.4} \ee
The two solutions are distinguished by the radius of $S^3$. In the
first case, the radius of $S^3$ is given by an arbitrary constant
$a<1$. It is possible to take the limit of small $a$ in which this
solution reduces to a flat space one. We will refer to this
$S^3$ as to the small $S^3$. In the second case, string is spinning
in the big $S^3$ of unit radius. In this case the flat space limit
does not exist. We will refer to this $S^3$ as to the big $S^3$. Now
we will discuss both solutions in turns.


\subsection{Strings Spinning in the Small $S^3$}


This solution is characterized by
\be
\sin \g =a <1\,.
\label{3.5} \ee
Equations for $\gamma$, $\psi$, $\v_1$ and $\v_2$ give
\be
\psi =n \sigma \,, \quad \v_1=\v_2 = n \tau\,.
\label{3.6} \ee
For simplicity, we will set $n=1$ and comment on the case $n \neq 1$
later. One can perform an $SO(4)$ rotation and
rewrite~\eqref{3.5}, \eqref{3.6} in the "chiral" form
\be
\sin \g=a\,, \quad \psi=\pi/4\,, \quad \v_1=\tau+\sigma\,, \quad
\v_2=\tau-\sigma\,.
\label{3.7} \ee
It is easy to check that in both cases $\sum_{i=1}^4 X_i^2 =a^2$ and,
thus, these two solutions are indeed related by an $SO(4)$ rotation.
Let us make a comment on the solution~\eqref{3.7}. If we substitute 
$\sin \gamma=a$, $\psi=\pi/4$ into the equations of motion we will 
find that they are satisfied if
\be
\partial_a \v_1 \partial^a \v_1=0 \,, \quad 
\partial_a \v_2 \partial^a \v_2=0\,.
\label{sol1}
\ee
This means that we could take both $\v_1$ and $\v_2$ to be 
a function of $\tau+\sigma$ or of $\tau-\sigma$. 
However, the second Virasoro condition~\eqref{Vir2}
requires that one of the angles, say $\v_1$, depend on $\tau+\sigma$
and the other one on $\tau-\sigma$.
This solution carries the equal angular momentum in both $\v_1$ and
$\v_2$ directions given by
\be
J_1=J_2 = J=\frac{\sqrt{\lambda}a^2}{2}\,.
\label{3.8} \ee
The Virasoro constraint~\eqref{Vir1} relates the 
$AdS_5$ and $S^5$ parts of the
solution. More precisely, it sets
\be
\kappa^2=2 a^2\,.
\label{3.9} \ee
Using eqs.~\eqref{3.4}, \eqref{3.8} and~\eqref{3.9} we obtain the
following relation between the energy and the angular momentum 
\be
E=\sqrt{4J \sqrt{\lambda}}\,.
\label{3.10} \ee
This solution has a limit of small $a$ in which the angular momentum
and the energy are very small and we get a string spinning in two
complex planes in flat space. On the other hand, since $a$ is
bounded from above the limit of infinite angular momentum does not
exist. This means that this state is dual to a short operator in
field theory. In particular, it was argued 
in~\cite{Bianchi, TT, RoibanT} that
some of these states are dual to 
members of the Konishi multiplet. See also~\cite{Liu}
for a worldsheet approach to studying spectra of short strings. 

Since this solution has a flat space limit, it is natural to expect
that the corresponding vertex operator has a structure similar to
that in flat space. Let us work in the coordinate system in which
the classical solution has the "chiral" form~\eqref{3.7}. Since
$\v_1$ carries only the "left moving" modes and $\v_2$ carries only
the "right moving" modes we write the vertex operator in the form
\be
V_J= e^{-E t_e}(\partial X)^J (\bar \partial Y)^J\,,
\label{3.11} \ee
where
\be
X= \sin \g \cos\psi e^{i \v_1}\,, \quad Y= \sin \g \sin\psi e^{i
\v_2}\,.
\label{3.12} \ee
In~\eqref{3.11}, for simplicity, we have set $\rho=0$.\footnote{It is
easy to check that the equation of motion for $\rho$ is satisfied
even when the vertex operators are inserted.} We want to compute the
correlation function
\be
\langle V_J (0) V_{-J}(\xi_1)\rangle \sim e^{-A_e}
\label{3.13} \ee
and to show that it produces the relation~\eqref{3.10}.

The Euclidean action $A_e$ is given by
\bea
A_e & = &\frac{\sqrt{\lambda}}{\pi} \int d^2 \xi( \partial t_e \bar
\partial t_e +\partial \g \bar \partial \g +\sin^2 \g \partial \psi
\bar \partial \psi +\sin^2 \g \cos^2 \psi \partial \v_1 \bar \partial
\v_1 +\sin^2 \g \sin^2 \psi \partial \v_2 \bar \partial \v_2 )
\nonumber \\
&+& E \int d^2 \xi (\delta^2 (\xi)- \delta^2 (\xi-\xi_1)) t_e
\nonumber \\
&-& J \int d^2 \xi \delta^2 (\xi) \ln (\partial X \bar \partial Y)
-J \int d^2 \xi \delta^2 (\xi-\xi_1) \ln (\partial \bar X \bar
\partial \bar Y)\,.
\label{3.14} \eea
As in the previous case, the expected semiclassical solution is
given by~\eqref{3.7} rewritten in the coordinates $(\xi, \bar \xi)$.
That is,
\bea
&& t_e=\frac{\kappa}{2} [\ln \xi \bar \xi -\ln (\xi-\xi_1)(\bar \xi-
\bar \xi_1)]\,,\nonumber \\
&& \sin \gamma=a\,, \quad \psi=\pi/4\,, \quad i \v_1= \ln \xi -\ln
(\xi-\xi_1)\,, \quad i \v_2 =\ln \bar \xi -\ln (\bar \xi-\bar
\xi_1)\,.
\label{3.15} \eea
Note that the holomorhic properties of $\v_1$ and $\v_2$ 
are encoded in the form of the vertex 
operator~\eqref{3.11}.
Let us first consider the equation of motion for $t_e$. It is the
same as in the previous section. We find that it is solved
by~\eqref{3.15} if $\kappa$ is related to energy as follows
\be
\kappa =\frac{E}{\sqrt{\lambda}}\,.
\label{3.16} \ee
Computing the corresponding part of the action gives
\be
A_e (t_e)=-\frac{E^2}{\sqrt{\lambda}} \ln |\xi|\,.
\label{3.17} \ee
Now we will consider the $S^3$ part of the action. It is
straightforward but tedious to show that with our ansatz~\eqref{3.15}
the equations of motion for $\gamma$ and $\psi$ are satisfied
provided
\be
a^2 =\frac{2 J}{\sqrt{\lambda}}\,.
\label{3.17.1} \ee
In proving this, we used the facts like
\begin{equation}
\frac{\delta^2 (\xi) X}{\partial X} =\frac{\delta^2 (\xi)}{\partial
\ln X}=\delta^2 (\xi) \xi (\xi-\xi_1)=0
\label{3.18} \ee
and similarly for $X$ replaced with $Y$ and $\delta^2 (\xi)$ replaced
with $\delta^2 (\xi- \xi_1)$. Now we consider the equation of
motion for $\v_1$. Using eq.~\eqref{3.18} it can be written as
\be
(\partial \bar \partial + \bar \partial \partial) i \v_1 = \frac{2
\pi J}{a^2 \sqrt{\lambda}}(\delta^2 (\xi)-\delta^2 (\xi-\bar
\xi_1))\,.
\label{3.20} \ee
From here we find the logarithmic solution for $\v_1$ provided
eq.~\eqref{3.17.1} is satisfied. One can check that the equation of
motion for $\v_2$ is analogously satisfied if eq.~\eqref{3.17.1} is
fulfilled.

Thus, we have found that~\eqref{3.17.1} is indeed a solution and the
parameters $\kappa$ and $a$ are fixed in terms of the quantum numbers
of the vertex operator $E$ and $J$. Now it is straightforward to
compute the $S^3$ part of the action $A_e$. Since $\gamma$ and
$\psi$ are constants and since $\v_1$ and $\v_2$ are
(anti)holomorphic the classical action  vanishes. The
delta-functional terms give
\be
A_e (S^3)= 4 J \ln |\xi_1|\,,
\label{3.22} \ee
where we ignored the obvious divergence $\sim \ln (0)$ and used the
relation between $a$ and $J$~\eqref{3.17.1}. Combining $A_e (S^3)$
and $A_e (t_e)$ we obtain
\be
\langle V_J (0) V_{-J}(\xi_1)\rangle \sim
|\xi_1|^{\frac{1}{\sqrt{\lambda}}(E^2- 4 J \sqrt{\lambda})}
\label{3.23} \ee
In the limit of large $E$ and $J$ we get the following
energy-angular momentum trajectory
\be
E=\sqrt{4 J \sqrt{\lambda}}\,
\label{3.24} \ee
which coincides with~\eqref{3.10}.
Note that eq.~\eqref{3.24} coincides with the flat space leading Regge
trajectory. The reason is that the angles $\gamma$ and $\psi$ are constants
and the consideration is essentially the same as in flat space.

We finish this subsection by noticing that one can generalize the
above analysis for the case $n\neq 1$. The corresponding vertex
operator in this case is given by
\be
V_J= e^{-E t_e} (\partial^n X)^J (\bar \partial^n Y)^J
\label{3.25}\ee
and the energy-angular momentum trajectory is
\be
E=\sqrt{4 J n \sqrt{\lambda}}\,.
\label{3.25.1} \ee
%


\subsection{Strings Spinning in the Big $S^3$}


The solution describing a string spinning in the big $S^3$ is characterized by %
\begin{equation}
\sin \gamma =1\,, \quad \gamma=\pi/2\,.
\label{3.26}
\end{equation}
Then the equations of motion give 
\begin{equation}
\psi=\pi/4\,, \quad \v_1=\omega \tau+ m \sigma\,, \quad \v_2 =
\omega \tau-m \sigma\,.
\label{3.27}
\end{equation}
Furthermore, from the Virasoro constraints it follows that 
\be
\kappa^2=\omega^2 +m^2\,.
\label{3.28}
\end{equation}
One can check that $\sum_{i=1}^4 X_i^4 =1$ so the string is spinning in 
$S^3$ or unit radius. Similarly to the case discussed in the previous 
subsection, the angular momenta in the $\v_1$ and $\v_2$ directions 
are still equal and are given by 
\be
J_1=J_2 =J=\frac{\sqrt{\lambda} \omega}{2}\,.
\label{3.29}
\end{equation}
From eqs.~\eqref{3.4}, \eqref{3.28} and~\eqref{3.29} we get the 
following relation between the energy and the angular momentum
\be
E=\sqrt{4 J^2 + m^2 \lambda}\,.
\label{3.30}
\end{equation}
Note that since the radius of $S^3$ is fixed to be unity
this solution does not have a flat space limit. On the other hand
it admits the large $J$ limit as well as the BMN~\cite{BMN} expansion
in powers of $\lambda/J^2$. 
Also note that, in addition to $E$ and $J$, this solution has one more 
quantum number. This is the winding number $m$. This winding number is 
topologically trivial and the solution is 
unstable~\cite{FTmulti, FTthree}. Similarly, we expect the 
corresponding quantum state to be unstable under decay to the BPS state 
with $m=0$ and $E=2 J$. 

Now we would like to discuss the vertex operator corresponding to 
this solution. Since this solution does not have a flat space limit 
the vertex operator must be more subtle than in the previous examples. 
Moreover, it has to depend not just on $E$ and $J$ but also on $m$. 
First, let us note that when $m=0$ we obtain a BPS state with $E=2 J$. 
The corresponding vertex operator then must be 
\be
V_{J, m=0}=e^{-E t_e}X^J Y^J\,,
\label{3.31}
\ee
where $X$ and $Y$ are given by eq.~\eqref{3.12} and just like 
in the previous subsection, for simplicity, we set $\rho=0$. 
Now we want to add a non-trivial $m$. Though it is topologically
trivial it still has a meaning of winding number. In flat space, 
to describe states with a non-trivial winding number one has to 
introduce the T-dual coordinates $\tilde{X}$ or write the vertex operators
in terms of the left- and right-moving fields and treat 
them as the independent 
variables. 
We would like to propose that a similar procedure 
should be performed in it this case. 
One more reason in favor of using the T-dual variables is
the following simple observation. The formula for the 
energy~\eqref{3.30} is invariant under the T-duality looking
transformation
\be
J \leftrightarrow \pm \frac{m \sqrt{\lambda}}{2}\,.
\label{3.31.1}
\ee
Thus, on general grounds, one can expect that it should be 
possible to perform calculations so that this symmetry 
is manifest. In particular, if this is the case,
it should be possible to write the vertex 
opearators in the form which respects this symmetry. 
T-duality in $AdS_5 \times S^5$ was studied 
in~\cite{Ricci1, BM, Ricci2}. 
However, in our case, as far as the semiclassical analysis is 
concerned, the discussion will be much simpler. 
First, we do not have to worry about the fermions. Second, since 
$\gamma$ and $\psi$ are constants, we have only the two 
decoupled field 
$\v_1$ and $\v_2$ whose actions are free. Thus, we 
can perform T-duality as
in flat space. Requiring that the vertex operator 
respects the symmetry~\eqref{3.31.1} we write it in the form
\be
V_{J, m}= e^{-E t_e} X^J Y^J e^{\frac{i \sqrt{\lambda}m}{2}\tilde{\v}_1}
e^{-\frac{i \sqrt{\lambda}m}{2}\tilde{\v}_2}\,,
\label{3.32}
\ee
where $\tilde{\v}_1$ and $\tilde{\v}_2$ are the T-dual field. 
We will see below that this expression is consistent with the 
classical solution~\eqref{3.27}. 
In particular, the coefficient in front of $\sigma$ will be exactly 
$m$ which is not obvious at this stage.
We would like to point out that our discussion in this subsection
is rather 
heuristic and that eq.~\eqref{3.32} is not the full vertex operator
but its simplified version. In principle, it can also depend on 
other $S^5$ coordinates. Fortunately, this is not relevant for our 
purposes since all the remaining coordinates are constants. 

As before, we consider 
$\langle V_{J, m}(0) V_{-J, -m}(\xi_1)\rangle \sim e^{-A_e}$, 
where the Euclidean action $A_e$ is given by
\bea
A_e & = &\frac{\sqrt{\lambda}}{\pi} \int d^2 \xi( \partial t_e \bar
\partial t_e +\partial \g \bar \partial \g +\sin^2 \g \partial \psi
\bar \partial \psi +\sin^2 \g \cos^2 \psi \partial \v_1 \bar \partial
\v_1 +\sin^2 \g \sin^2 \psi \partial \v_2 \bar \partial \v_2 )
\nonumber \\
&+& E \int d^2 \xi (\delta^2 (\xi)- \delta^2 (\xi-\xi_1)) t_e 
- J \int d^2 \xi \delta (\xi) \ln (X Y)
-J \int d^2 \xi 
\delta (\xi-\xi_1) \ln (\bar X \bar Y)
\nonumber \\
&-& \frac{i \sqrt{\lambda}m}{2} \int d^2 \xi 
(\delta^2 (\xi)- \delta^2 (\xi-\xi_1))(\tilde{\v}_1-\tilde{\v}_2)\,.
\label{3.32.1}
\eea
The part of the action depending on $t_e$ is identical to the previous cases.
It leads to the solution
\be
t_e=\frac{E}{2 \sqrt{\lambda}} [\ln \xi \bar \xi - \ln (\xi-\xi_1)
(\bar \xi-\bar \xi_1)]
\label{3.33}
\ee
and to the contribution to the action given by
\be
A_e (t_e) =-\frac{E^2}{\sqrt{\lambda}} \ln |\xi_1|\,.
\label{3.34}
\ee
Furthermore, the equations for $\gamma$ and $\psi$ are trivially 
satisfied since $\cos \gamma=0$ and 
$\sin \psi=\cos \psi$.\footnote{If the winding part of the 
vertex operators depends 
on $\gamma$ and $\psi$ there is an additional 
contribution to the singular part of the equations of motion. 
Here we are assuming that they are still satisfied if $\g=\pi/2$, 
$\psi=\pi/4$.} 
Thus, we can concentrate on the action and the operators depending on 
$\v_1$ and $\v_2$ only. The relevant classical action is 
\be
A_e (\v_1, \v_2)=\frac{\sqrt{\lambda}}{2\pi} \int d^2 \xi
(\partial \v_1 \bar \partial \v_1+ \partial \v_2 \bar \partial \v_2)
\label{3.35}
\ee
and the relevant vertex operator is 
\be
V_{J, m}=e^{i J \v_1}e^{\frac{i \sqrt{\lambda}m}{2} \tilde{\v}_1}      
e^{i J \v_2} e^{\frac{-i \sqrt{\lambda}m}{2} \tilde{\v}_2}\,. 
\label{3.36}
\ee
Clearly, it is enough to consider only one field, say $\v_1$, since 
the contribution from the second one is obtained by replacing 
$m \to -m$. The consideration will be relatively standard.
Nevertheless, for completeness we will perform it below. 
First, we will write the action using both $\v_1$ and $\tilde{\v}_1$.
It reads
\be
A_e(\v_1, \tilde{\v}_1)=\frac{\sqrt{\lambda}}{8 \pi} \int d^2 \xi
[ -(\partial-\bar \partial)\v_1  (\partial-\bar \partial)\v_1 -
(\partial-\bar \partial)\tilde{\v}_1  
(\partial-\bar \partial)\tilde{\v}_1
+ 2 (\partial+\bar \partial)\v_1 (\partial-\bar \partial)\tilde{\v}_1]\,.
\label{3.37}
\ee
If we find the equation of motion for $\tilde{\v}_1$
and substitute it back in~\eqref{3.37} we obtain the action 
for $\v_1$. 
The equations of motion for $\v_1$ and $\tilde{\v}_1$ are as 
follows~\footnote{Note that the equations of motion by construction must be invariant under
$\v_1 \leftrightarrow \tilde{\v}_1$, $J \leftrightarrow m \sqrt{\lambda}/2$.}
\bea 
&&(\partial  -\bar \partial )^2 \v_1
-(\partial +\bar \partial) 
(\partial  -\bar \partial) \tilde{\v}_1=
\frac{4 \pi i J}{\sqrt{\lambda}} (\delta^2 (\xi)-\delta^2 (\xi-\xi_1))\,,
\label{2.37.1} \\
&&(\partial  -\bar \partial )^2 \tilde{\v}_1
-(\partial +\bar \partial) 
(\partial  -\bar \partial) \v_1=
2 \pi i m (\delta^2 (\xi)-\delta^2 (\xi-\xi_1))\,.
\label{2.37.2}
\eea
It is straightforward to check that these equations are solved by 
\be
i \v_1 = \frac{J}{\sqrt{\lambda}} [\ln \xi \bar \xi -
\ln (\xi-\xi_1) (\bar\xi -\bar \xi_1)]+
\frac{m}{2} [\ln  \xi /\bar \xi -
\ln (\xi-\xi_1) /(\bar\xi -\bar \xi_1)]
\label{2.37.3}
\ee
and
\be
i \tilde{\v}_1 = \frac{m}{2} [\ln \xi \bar \xi -
\ln (\xi-\xi_1) (\bar\xi -\bar \xi_1)]+
\frac{J}{\sqrt{\lambda}} [\ln  \xi /\bar \xi -
\ln (\xi-\xi_1) /(\bar\xi -\bar \xi_1)]\,.
\label{2.37.4}
\ee
If we recall the conformal transformations~\eqref{1.1} and~\eqref{1.2}
and rewrite $\v_1$ in terms of $\tau$ and $\sigma$ we will get 
precisely the classical solution 
\be
\v_1 = -i \omega \tau_e + m \sigma\,, 
\label{2.37.5}
\end{equation}
where 
\be 
\omega=\frac{2 J}{\sqrt{\lambda}}
\label{2.37.6}
\ee
and the coefficient in front of sigma is precisely $m$. Note that 
we wrote 
our vertex operator imposing the symmetry~\eqref{3.31.1}. 
However, it turned out to be what is needed to reproduce
the classical solution. This can be viewed as a consistency check 
on our proposed vertex operator. 

Evaluating the action on our solution we obtain
\be 
A_e (\v_1)= \frac{1}{\sqrt{\lambda}}
\left(J+\frac{\sqrt{\lambda}m}{2}\right)^2
\ln \xi_1 + \frac{1}{\sqrt{\lambda}}
\left(J-\frac{\sqrt{\lambda}m}{2}\right)^2
\ln \bar \xi_1\,.
\label{3.48}
\ee
Note that if $m \neq 0$ the coefficients at $\ln \xi_1$ and 
$\ln \bar \xi_1$ are different. This means that $e^{-A_e (\v_1)}$
does not behave as $|\xi_1|$ to some power which is not 
consistent with conformal symmetry. To achieve conformal 
symmetry we need 
to add the contribution from
the second 
field $\v_2$. 
Its action is the same as above but with $m \to -m$. 
Then the action of these two fields is
\be
A_e (\v_1, \v_2)= A_e (\v_1)+A_e (\v_2)=
\frac{4 J^2 +m^2 \lambda}{\sqrt{\lambda}} \ln |\xi_1|\,.
\label{3.49}
\ee
Combining this with $A_e(t_e)$ we get 
\be
\langle V_{J, m}(0) V_{-J, -m}(\xi_1)\rangle\sim
|\xi_1|^{\frac{1}{\sqrt{\lambda}} (E^2- 4 J^2 -m^2 \lambda)}\,, 
\label{3.50}
\ee
from which we obtain the dependence of the energy 
on $J$ and $m$ of the form
\be
E=\sqrt{4 J^2 +m^2 \lambda}\,.
\label{3.51}
\ee
This coincides with eq.~\eqref{3.30} and we have an agreement 
between the proposed semiclassical vertex operator~\eqref{3.32}
and the classical string solution~\eqref{3.26}, \eqref{3.27}.



\section*{Acknowledgements}


The author would like to thank Arkady Tseytlin for
many valuable discussions and comments on the draft.
The work is supported by an STFC Fellowship.




\begin{thebibliography}{99}

\bibitem{BRP}
I. Bena, J. Polchinski and R. Roiban, 
``Hidden Symmetries of the $AdS_5 \times S^5$ Superstring,''
Phys.Rev.D69:046002,2004 [arXiv:hep-th/0305116].

\bibitem{AF}
G. Arutyunov and S. Frolov, 
``Foundations of the $AdS_5 \times S^5$ Superstring. Part I,''
arXiv:0901.4937 [hep-th].

\bibitem{Juan}
J. M. Maldacena, 
"The Large $N$ Limit of Superconformal Field Theories and Supergravity,"
Adv.Theor.Math.Phys.2:231-252,1998 [arXiv:hep-th/9711200].

\bibitem{Gubser}
S. S. Gubser, I. R. Klebanov and A. M. Polyakov, 
"Gauge Theory Correlators from Non-Critical String Theory,"
Phys.Lett.B428:105-114,1998 [arXiv:hep-th/9802109].

\bibitem{Witten}
E. Witten, 
"Anti De Sitter Space And Holography,"
Adv.Theor.Math.Phys.2:253-291,1998 [arXiv:hep-th/9802150].

\bibitem{Beisert1}
N. Beisert and M. Staudacher, 
"Long-Range $PSU(2,2|4)$ Bethe Ansaetze for Gauge Theory and Strings,"
Nucl.Phys.B727:1-62,2005 [arXiv:hep-th/0504190]. 

\bibitem{Beisert2}
N. Beisert,
"The $su(2|2)$ Dynamic S-Matrix,"
Adv.Theor.Math.Phys.12:945,2008 [arXiv:hep-th/0511082].

\bibitem{G1}
N. Gromov, V. Kazakov and P. Vieira,
"Integrability for the Full Spectrum of Planar AdS/CFT,"
Phys.Rev.Lett.103:131601,2009 [arXiv:0901.3753 [hep-th]].

\bibitem{F}
D. Bombardelli, D. Fioravanti and R. Tateo,
"Thermodynamic Bethe Ansatz for planar AdS/CFT: a proposal,"
J.Phys.A42:375401,2009 [arXiv:0902.3930 [hep-th]].

\bibitem{G2}
N. Gromov, V. Kazakov, A. Kozak and P. Vieira,
"Integrability for the Full Spectrum of Planar AdS/CFT II,"
arXiv:0902.4458 [hep-th] 
 
\bibitem{AFint}
G. Arutyunov and S. Frolov, 
``Thermodynamic Bethe Ansatz for the 
$AdS5 \times S5$ Mirror Model,'' 
JHEP 0905, 068 (2009) [arXiv:0903.0141 [hep-th]].

\bibitem{Polyakov}
A. M. Polyakov, 
"Gauge Fields and Space-Time,"
Int.J.Mod.Phys. A17S1 (2002) 119-136 [arXiv:hep-th/0110196].   

\bibitem{Tseytlinver}
A. A. Tseytlin, 
``On semiclassical approximation and spinning string vertex operators in $AdS_5 \times S^5$,''
Nucl.Phys.B664:247-275,2003 [arXiv:hep-th/0304139].

\bibitem{Tseytlinrev}
A.A. Tseytlin, ``Spinning strings and AdS/CFT duality,''
``From Fields to Strings: Circumnavigating Theoretical Physics'', 
M. Shifman, A. Vainshtein, and J. Wheater, eds. (World Scientific, 2004)
[arXiv:hep-th/0311139]. 

\bibitem{GKP}
S. S. Gubser, I. R. Klebanov and A. M. Polyakov,
``A semi-classical limit of the gauge/string correspondence,''
Nucl.Phys. B636 (2002) 99-114 [arXiv:hep-th/0204051]

\bibitem{Ricci1}
R. Ricci, A. A. Tseytlin and M. Wolf, 
``On T-Duality and Integrability for Strings on $AdS$ Backgrounds,''
JHEP 0712:082,2007 [arXiv:0711.0707 [hep-th]].

\bibitem{BM}
N. Berkovits and J. Maldacena,
``Fermionic T-Duality, Dual Superconformal Symmetry, and the Amplitude/Wilson Loop Connection,''
JHEP0809:062,2008 [arXiv:0807.3196 [hep-th]].

\bibitem{Ricci2}
N. Beisert, R. Ricci, A. A. Tseytlin and M. Wolf, 
``Dual Superconformal Symmetry from $AdS_5 \times S^5$ Superstring Integrability,''
Phys.Rev.D78:126004,2008 [arXiv:0807.3228 [hep-th]].

\bibitem{FTsemi}
S. Frolov and A. A. Tseytlin,
``Semiclassical quantization of rotating superstring in $AdS_5 \times S^5$,''
JHEP 0206:007,2002 [arXiv:hep-th/0204226].

\bibitem{FTmulti}
S. Frolov and A. A. Tseytlin, 
``Multi-spin string solutions in $AdS_5 \times S^5$,''
Nucl.Phys.B668:77-110,2003 [arXiv:hep-th/0304255].
 
\bibitem{Bianchi}
M. Bianchi, J. F. Morales and H, Samtleben,
"On stringy $AdS_5 \times S^5$ and higher spin holography 
JHEP 0307 (2003) 062 [arXiv:hep-th/0305052].

\bibitem{TT}
A. Tirziu, A.A. Tseytlin,
"Quantum corrections to energy of short spinning string in $AdS5$,"
Phys.Rev.D78:066002,2008 [arXiv:0806.4758 [hep-th]].

\bibitem{RoibanT}
R. Roiban and A. A. Tseytlin,
``Quantum strings in $AdS_5 \times S^5$: 
strong-coupling corrections to dimension of Konishi operator,''
JHEP 0911:013,2009 [arXiv:0906.4294].

\bibitem{Liu}
B. A. Burrington and J. T. Liu,
``Spinning strings in $AdS_5 \times S^5$: A worldsheet perspective,''
Nucl.Phys.B742:230-252,2006 [arXiv:hep-th/0512151]. 

\bibitem{BMN}
D. Berenstein, J. Maldacena and H. Nastase,
``Strings in flat space and pp waves from ${\cal N}=4$ Super Yang Mills,''
JHEP 0204 (2002) 013 [arXiv:hep-th/0202021].

\bibitem{FTthree}
S. Frolov and A. A. Tseytlin, 
``Quantizing three-spin string solution in $AdS_5 \times S^5$,''
JHEP 0307:016,2003 [arXiv:hep-th/0306130].


\end{thebibliography}
\end{document}